\documentclass{article}
\usepackage{sbc-template-Custom}

\usepackage[utf8]{inputenc} 
\usepackage[T1]{fontenc}    
\usepackage{hyperref}       
\usepackage{url}            
\usepackage{booktabs}       
\usepackage{amsfonts}       
\usepackage{nicefrac}       
\usepackage{microtype}      
\usepackage{lipsum}
\usepackage{amsmath}
\usepackage{graphicx}
\usepackage{geometry}

\title{Hit by the Data: a visual data analysis regarding the effects of traffic public policies}
\authors{Luana Müller; Camila Moser; Guilherme Paris;
    \nextauthors Lucas Freitas; Mayara Oliveira; Wagner Signoretti;
    \nextauthors Isabel Manssour; Milene Silveira}
\address{
    School of Technology
    \nextinstitute Pontifical Catholic University of Rio Grande do Sul
    \nextinstitute Porto Alegre, Brazil
}
\emails{
     \texttt{luana.muller@acad.pucrs.br; camila.moser@acad.pucrs.br; guilherme.paris@acad.pucrs.br}
     \nextemails \texttt{lucas.freitas.003@acad.pucrs.br; mayara.jesus@acad.pucrs.br wagner.signoretti@acad.pucrs.br}
     \nextemails \texttt{manssour@pucrs.br; milene.silveira@pucrs.br} }

\begin{document}
\maketitle

\begin{abstract}
The availability of Open Government Data (OGD) provides means for citizens to understand and follow governmental policies and decisions, showing evidence of how the latter have contributed to both the place they live in and their lives. In such a scenario, one of the proposals is the use of visualizations to support the process of data analysis and interpretation. Herein, we present the use of three different visualization tools - a commercial one and two academic ones – applied to two specific Brazilian cases: the implementation of the Drink Driving Law and the construction of a new overpass in an important city avenue. Our focus was on the analysis of how visualization could help in the identification of the effects of such traffic public policies. As our main contributions, we present details on the effects of the observed policies, as well as new cases showing how visualization tools can assist users to interpret OGD.
\end{abstract}

\small{}

\section{Introduction}

The increasing availability of Open Government Data (OGD), allied to the lack of standardization in the way they are presented, leads to great difficulty when it comes to their use by citizens in general. Every citizen who makes professional use of these kinds of data needs means to analyze and get insight about them. In this scenario, Graves~\cite{Graves2013} advocates the use of ``visualizations as a medium to consume, share and interact with data.'' According to Sivarajah et al.~\cite{Sivarajah2016}, ``the use of visualization techniques makes the data analysis task easier for human interpretation and provides support to decision making''.

We can find several investigations exploring the use of visualization to support OGD analysis~\cite{Graves2015, Santos2018}. In the Brazilian context, for instance, the work from De Mendonça, Maciel and Filho~\cite{deMendonca2014} presents a case study in which a map is created to visualize information regarding the infestation of Aedes aegypti mosquito in the city of Cuiabá. Still in the Brazilian context, Craveiro and Martano~\cite{Craveiro2015} reported that, despite the data availability on government portals, these data have not always been understood by the broader public. Thus, the authors presented a tool called \textit{Cuidando do Meu Bairro} (Looking After My Neighborhood), used in São Paulo city to promote a better visualization of government spending, and, therefore, to foster citizen engagement. 

Despite different attempts to create visualization tools that would help in better understanding OGD, the systematic mapping on OGD visualization presented by Eberhard and Silveira~\cite{Eberhardt2018} shows us that one of the major challenges in this field refers to the skills required to use such tools~\cite{Brugger2016,Graves2015,Pirozzi2016}. 

In this context, Méndez, Hinrichs and Nacenta~\cite{Mendez2017} performed a study in which they compared people's visualization processes using two visualization tools (Tableau Desktop and iVoLVER) and discussed how different approaches can influence the visualization process, the decisions on the visualization design, the sense of control and authorship, and the enthusiasm to explore alternative designs. 

In the research herein presented, we also chose to investigate different visualization tools: a commercial one and two academic ones. In our case, the investigation was based on the analysis of OGD visualizations applied to two specific Brazilian cases: the implementation of the Drink Driving Law and the construction of a new overpass. We analyzed the data related to the year before the events had taken place (the law implementation and the overpass construction) and 2016, the last year we have associated with the available OGD. Our focus was the analysis of how visualization could assist in the identification of the effects of such traffic public policies. 

The remainder of this paper is organized as follows. Section~\ref{sec:method} describes our research methodology, including the used dataset, the selected case studies, and the used visualization tools. The results of our analysis with our findings are presented in Section~\ref{sec:Findings}, to be further discussed in Section~\ref{sec:Discussion}. Finally, we end this paper with our conclusions and future work directions in Section~\ref{sec:conclusions}.

\section{Research Methodology}
\label{sec:method}

The research methodology followed in this work has been split into five steps: I) literature review; II) selection of a dataset; III) identification of case studies; IV) selection of three different tools; V) performance of visual analysis and finding identification. Moreover, the following subsections present an explanation on the used data, a description of the selected case studies, and, lastly, a presentation of the visualization tools used in our visual analysis. Our findings will be discussed in depth in the next section.

\subsection{Used Dataset}

For our visual analysis, we used OGD related to traffic accidents that had occurred in a city in southern Brazil, called Porto Alegre. An open government data portal~\footnote{http://datapoa.com.br/dataset/acidentes-de-transito}, maintained by the City Hall, records OGD about traffic accidents that occurred between 2000 and 2016, in csv format. These records generate large multivariate datasets, with several attributes associated with a specific location (latitude and longitude), such as accident type, time, date, number of injuries, fatalities, vehicle category, and weather conditions.

However, as these data have different format depending on the year, we chose the data referring to the year of 2016 to be used as a model for the standardization of other years. We also removed some attributes that were unnecessary for our analysis, because there were either data we did not consider as useful (e.g. consortium), or data with the same information, or even because some data were blank in most records. Besides that, we had to deal with missing data and non-standardized data format, which was the case of the date. For these standardizations, we have used OpenRefine~\footnote{openrefine.org}.

\subsection{Case Studies}

We chose two case studies to analyze how visualization could help in the identification of the effects of traffic public policies. Our main goal is to identify how laws and street works have helped in reducing traffic accidents in the city.

The first case study is related to the Drink Driving Law
~\footnote{Law number 11.705 - https://bit.ly/2v2sXBy}, which was enacted on June 19, 2018. This law prohibits the consumption of alcohol by drivers, imposing more severe penalties for those who drive under the influence of alcohol. Its main purpose is to avoid accidents that may occur due to the carelessness of drivers with impaired operating abilities. Thus, we decided to evaluate the total number of accidents that happen on weekends (from Friday to Sunday), when people in Brazil usually go to parties, bars and restaurants, that is, places where there is typically more consumption of alcoholic beverages. It was mainly on weekends that "Balada Segura"~\footnote{https://baladasegura.rs.gov.br/inicial} also began to be held in 2012. The term, freely translated into English as "Safe Clubbing", consists of a surveillance operation carried out by traffic organs, Military and Civil Police, which aims at preventing alcohol-related traffic criminal offences in places and at times with a higher incidence of accidents. Considering these dates, we opted for analyzing the total number of accidents in 2007, before the Drink Driving Law, and in 2016, after people had already been aware of the zero tolerance laws and drunk driving inspections.

The construction of an overpass on an important avenue of Porto Alegre, Brazil, was the second chosen case study. The idea was to evaluate how much the construction of the overpass has influenced the reduction of traffic accidents once it was a very busy crossroad. Since the overpass construction began in August 2012, being only launched in June 2015, we chose to analyze the total number of accidents in 2011 and 2016, before and after its construction.

\subsection{Visualization Tools}

Three tools were chosen to our proposed analysis: a commercial and two academic ones (developed in our research group). The next subsections describe them in details.

\subsubsection{Tableau}

It is a commercial software for data analysis and visualization, both for individual and group analysis~\footnote{https://www.tableau.com}. It comprises several tools required to generate visualizations, from receiving data files of various formats to the generation of different charts and dashboards. The provided interface allows a quick selection and optimization of data, as well as suggesting chart models for visualization, such as treemaps, bars, bubble, pie, line, scatter, among others.  

\subsubsection{GeoCharts}

This tool presents an interactive visualization design for visual analysis of geospatial multivariate data that facilitates data analysis and knowledge discovery. It allows the representation of several attributes on the map with dynamically linked charts and an interactive approach based on the brushing and linking technique integrated with coordinated multiple views that can support visual analysis. Thus, a set of interactive visualizations are combined to stimulate an active interaction of the end user. It offers a way to interactively apply different filters and analyze several attributes, updating all the related visual representations, including the map with clustered charts.

\subsubsection{Traffic Accidents Analyzer (TrAcc)}
This tool provides different visualization techniques specifically for the OGD of traffic accidents. It provides an animated heat map based on a timeline, as well as three charts that help visualize the information contained in the data: a Historical Bar Chart, which shows the number of accidents within a time range (0h to 23h); a Pie Chart, showing the regions of the city (north, south, east, and center) where most accidents occur; and a Horizontal Bar Chart, showing the types of accidents and their total number. Several filter options are also available, such as the type of vehicle (car, motorcycle, truck, etc.), time range (morning, afternoon, night, dawn), weather condition (good, rainy, cloudy), and days of the week.

\section{Findings}
\label{sec:Findings}

As previously mentioned, we chose two scenarios to support our case studies. A common step to both case studies was loading the data into the tools, which enabled us, after the import completion, to perform the visualizations.

Regarding the steps described in Subsection 2.1, it was fairly simple to load the data into GeoCharts, since it allows the use of any data in the csv format. Tableau also supports loading csv data. However, some problems happened when the files were opened, not creating the columns correctly, for instance. Nonetheless, these problems were easily solved by just converting the files to xls format. One situation identified in both tools, GeoCharts and Tableau, was related to the latitude and longitude, which, in some cases, were not readable by them. The solution to this issue was just to change the decimal separator format. TrAcc did not present any problems, once the data were already embedded in it.

The subsections below describe the findings grouped by each case study, also presenting the visualizations which helped us throughout the analysis.  

\subsection{Case Study: Drink Driving Law}

\begin{figure*}[h]
\centering
\includegraphics[scale=0.6]{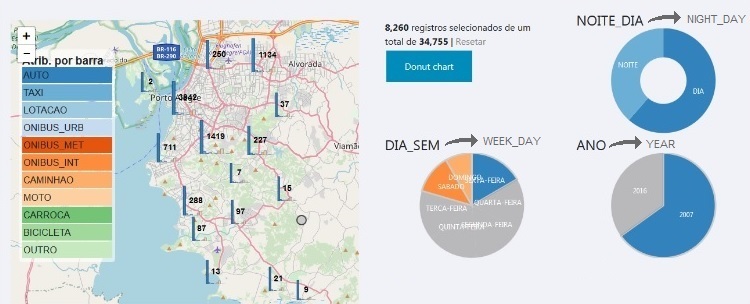}
\caption{General overview regarding traffic accidents in 2007, by GeoCharts}
\label{fig:GeoChartsLeiSeca2007}
\end{figure*}

In our first case study on the effects of the Drink Driving Law, we started analyzing the Geocharts tool. To reach our goal, it was necessary to apply three filters while using GeoCharts: year, day of the week (Friday, Saturday, and Sunday), and shift. After that, GeoCharts showed a map and different associated charts presenting the numbers of traffic accidents in 2007 and 2016 (Figure ~\ref{fig:GeoChartsLeiSeca2007}~\footnote{The used systems were developed in Portuguese. We provide English subtitles in the figures.}). In order to filter data by a specific year, users must click and select the year they want in the year chart. We highlight that, regarding these charts, users could pre-select the variables they wanted to visualize in charts along the map (five charts at most). At any time, in case users wanted to change the charts, they only needed to select new variables and generate a new visualization.

In general terms, the results showed that in 2007, before the implementation of the law, 5076 accidents had occurred  during the day and 3180 at night, whereas in the year of 2016 the number of accidents had reduced to 3083 occurrences during the day and 1345 during the night. In GeoCharts, the number of occurrences is numerically presented next to the map at the top, in the same figure area as the charts.

 By using Tableau and TrAcc to achieve the same results, it was necessary to apply the same three filters. Regarding these filters, TrAcc offers six filtering options: by vehicle type, day shift, weather condition, day of the week, address, and year. Tableau, on the other hand, offers the possibility of choosing any variable from the dataset to be used as a filter.
 
\begin{figure}
\centering
\includegraphics[scale=0.4]{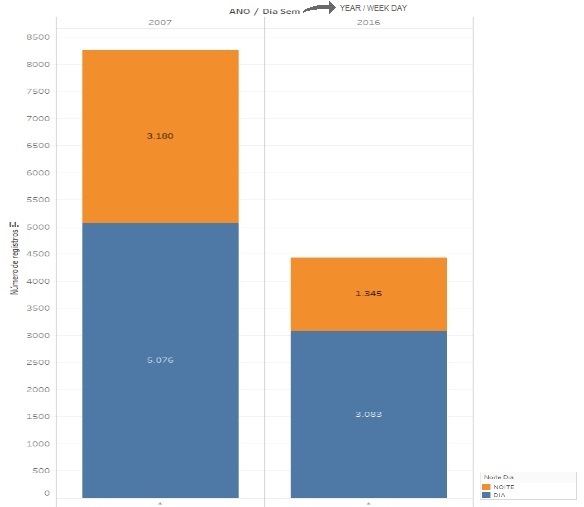}
\caption{General overview regarding traffic accidents in 2007 and 2016, by Tableau}
\label{fig:TableauLeiSeca2007}
\end{figure}

 Considering forms of visualization presentation, such as GeoCharts, Tableau allows us to jointly visualize the data from different time periods. However, it only presents one chart at a time, and users must switch among different types of charts (for instance, in Figure ~\ref{fig:TableauLeiSeca2007} we present the bar chart, chosen to present a simple overview regarding the general numbers from both years).

\begin{figure}[h]
\centering
\includegraphics[scale=0.6]{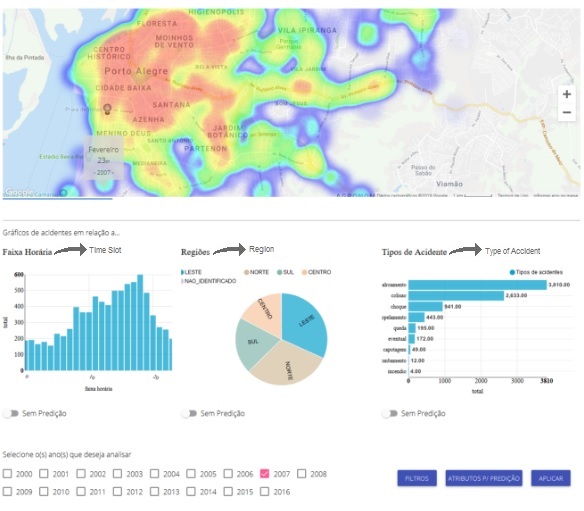}
\caption{General overview regarding traffic accidents in 2007, by TrAcc}
\label{fig:TraccLeiSeca2007}
\end{figure}

Regarding the visualization provided by TrAcc, it shows a heat map in which it is possible to visualize the data from the selected years (Figure ~\ref{fig:TraccLeiSeca2007}). As compared to GeoCharts, TrAcc also provides different charts that help bring signification to the data. Differently from the other tools, in which it is simple to identify the totals, the information in TrAcc is presented in a partitioned way through the charts. To verify the totals, it is necessary to manually calculate the information by summing up the data presented in the chart \textit{Type of Accident}, in which the numbers are clearly presented.

\subsection{Case Study: Overpass construction}

In regard to the second case study, we analyzed the impact of the construction of an overpass in an important city avenue in southern Brazil. 

\begin{figure}[h]
\centering
\includegraphics[scale=0.6]{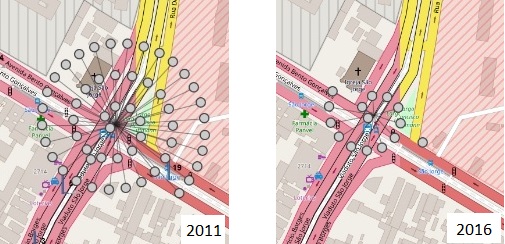}
\caption{Visualization of accidents on the crossroad in 2011 and 2016, by GeoCharts}
\label{fig:GeoChartsBento2011_2016}
\end{figure}

By using GeoCharts, it was required to filter by year, and to zoom in and find the desired location on the map. After finding the location, users need to analyze and check the number presented on the map. The number we found is related to accidents which had occurred around the crossroad area, not exactly in the intersection of the two avenues. We identified 53 accidents in 2011 on the crossroad where the overpass is located nowadays (Figure ~\ref{fig:GeoChartsBento2011_2016}  left). Following the same steps, we verified that the number of accidents in the same region was reduced to 14 in 2016. 

\begin{figure}[h]
\centering
\includegraphics[scale=0.5]{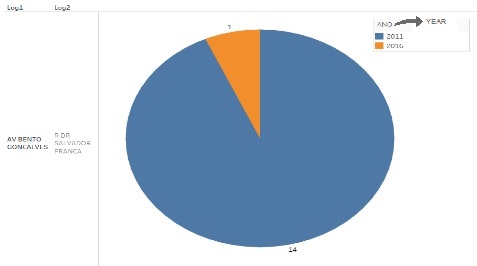}
\caption{Visualization of accidents in 2011 and 2016 on the crossroad, by Tableau}
\label{fig:TableauBento20112016}
\end{figure}

By using Tableau, we applied a filter regarding the years, and also filters regarding the required address (in this case, we informed two addresses, representing the crossroad). Thus, differently from GeoCharts, the tool provided us with a chart visualization about the accidents that had taken place exactly on the intersection of the avenues. By that, we found that, in the year of 2011, 14 accidents had occurred  in this crossroad, and, in 2016, after the construction of the overpass, the number reduced to only one accident (Figure ~\ref{fig:TableauBento20112016}).

Similar to Tableau, TrAcc provides a filter in which it is possible to inform the year (as presented in Figure ~\ref{fig:TraccLeiSeca2007}), and filters users can apply to determine the addresses they want to analyze. The results presented by TrAcc in the heat map are not influenced by the addresses. However, the associated charts change, presenting the numbers related only to the informed addresses.

\section{Discussion}
\label{sec:Discussion}

The tools we investigated present different visualizations, helping us analyze the available data under different perspectives. 

In the next subsections, we firstly discuss aspects related to specific possibilities of use of the analyzed tools, followed by how these possibilities helped us during our case study analysis.

\subsection{Visualization strengths}

Regarding \textbf{data import}, TrAcc displays the data related to accidents from Porto Alegre city already loaded in the system, which prevents the user from having to deal with details regarding pre-processing and data import. However, Tableau and GeoCharts accept any type of dataset and can be used for different purposes. Also, after the data import, it is very simple for the user to change the charts that are being presented (GeoCharts allows the presentation of five charts at the same time, as well as allowing the user to choose five visulization variables in these charts), while in TrAcc the three charts and six filtering variables provided by the tool (besides the heat map) cannot be changed.

In regard to \textbf{filtering precision}, in the overpass case study, it calls our attention that GeoCharts, differently from Tableau and TrAcc, did not allow us to inform the addresses we wanted to visualize and, consequently, did not allow us to precisely check for accidents in a particular location. On the other hand, by using the zoom, GeoCharts provides a macro view, as presented in Figure ~\ref{fig:GeoChartsBento2011_2016} (right), that allows us to have a broader perspective of the analysis. By using the other tools, the users would need to know the exact description of surrounding addresses in order to visualize them, although a single accident occurred at the exact crossroad address (as presented by the other tools), there were 14 accidents in the area.  

\begin{figure}[h]
\centering
\includegraphics[scale=0.5]{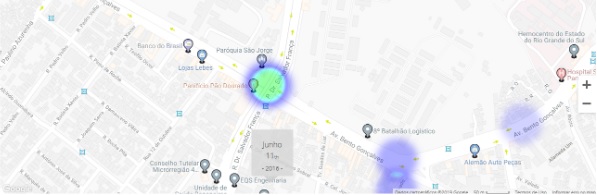}
\caption{Visualization of accidents in 2016 around the crossroad area, by TrAcc}
\label{fig:TrAccBento2016Zoom}
\end{figure}

Also, GeoCharts allows us to click in each one of the bullets presented in the image, providing \textbf{detailed information} about the accident (such as the kind of vehicle, weather conditions, day of the week, etc.). TrAcc, in turn, also allows zooming in and out the map to approach a location (Figure ~\ref{fig:TrAccBento2016Zoom}). However, it does not allow to get more information regarding the accidents. Despite the lack of information, the user has a fast overview of the accidents that had happened in that area through the heat visualization, whereas only numbers are presented in GeoCharts.

Regarding the possibility of \textbf{comparing different periods} in the same chart, Tableau allows it to be done, while this is not possible by using the other tools. In both GeoCharts and TrAcc the user needs to check the periods one at the time. The user can select different years to create the visualization, but the results are presented based on a concatenation of the data from the selected years. As presented in Figure ~\ref{fig:GeoChartsLeiSeca2007}, there is a chart (located in the figure's bottom right corner) with information regarding the years 2007 and 2016, and the user must click on the period he wants to visualize with more details. The presentation of the results in Tableau makes the visualization easier, especially comparing different periods, as presented in Figure ~\ref{fig:TableauBento20112016}, but it does not offer more details about those years. If the user wants to explore such information, he needs to generate a new visualization, choosing the variables he wants to visualize.

\subsection{Interpretation support}

In regard to how these different visualizations assist the interpretation of data, we would like to discuss and focus on each case study's needs.

The analysis of the Drink Driving Law effects requires an \textbf{overview} regarding the situation in the whole city. To achieve a simple and straightforward analysis, the visualization from Tableau may be enough, once it shows the numbers and charts related to it. Since we are dealing with information from locations all over the city, a visualization with many details could generate a negative effect on the users, since it would take a lot of effort to interpret this view. However, in need of a detailed perspective, both GeoCharts and TrAcc allow the users to identify the areas from the city in which the new law impacted more, and this insight is even simpler by the visualization through the heat map offered by TrAcc. Furthermore, the charts from GeoCharts allow fast filtering and the visualization of numbers, providing a quick response and helping users while they are interacting and interpreting these data.

About the second case study, we wanted to visualize a situation encompassing a very specific part of the city (a crossroad) in order to understand if the construction of an overpass had effectively helped in traffic. Unlike the first case study, in this one \textbf{details} can improve interpretation and are easy to be visualized (since our analysis involved a very small area of the city). Regarding the precise view required in this case, both Tableau and TrAcc allow the users to inform the address they want to visualize (in this case, two addresses corresponding to the intersection of two avenues). However, this resource can lead users to some false interpretation, once the accidents that had happened around the crossroad are not presented to them. GeoCharts, on the other hand, does not allow users to inform the address to be visualized; therefore, users need to zoom in to find the location they want (a task that can become costly in cases where the city is large and the user is not familiar with the city map). Thus, the possibility to inform the address must be available to users; however, it cannot limit the visualization without presenting the areas around it. Also, to provide a richer interpretation, once the scenario is limited to a small part of the city, details regarding the accidents must support the user, like those presented by GeoCharts and TrAcc.

\section{Conclusion and Future Work}
\label{sec:conclusions}

Nowadays, it is easy to find Open Government Data (OGD) available on governmental portals, but it is not so easy to employ them without some help. Similar to other researchers in the field, we advocate that the use of visualization techniques makes it easier for citizens to analyze and interpret these kinds of data. 

In the research herein presented, we used three different visualization tools to analyze the OGD referring to two Brazilian cases: the implementation of the Drink Driving Law in 2007 and the construction of a new overpass in an important crossroad, which started in 2011 and was concluded in 2016. We focused on analyzing how visualizations could support the identification of the effects of the studied traffic public policies, and we presented a twofold contribution: first, the details on the effects of the observed policies, and, second, two more cases showing how visualization tools could help users to interpret OGD.

As for our next research steps, we intend to analyze the same cases with professionals who closely work with traffic and traffic policies to gather their opinion about the tools' usage and their possibilities. We also intend, as future work, to analyze new case studies, enabling us to explore other perspectives of those given tools.

\bibliographystyle{unsrt}  
\bibliography{references} 

\end{document}